\documentstyle[12pt,preprint,aps,epsf]{revtex}

\tighten

\begin{document}
\draft
\title{ Application of a semi-microscopic core-particle coupling method
to the backbending in odd deformed nuclei} 
\author{
Pavlos Protopapas
\footnote{ pavlos@walet.physics.upenn.edu}
and Abraham Klein
\footnote{ aklein@walet.physics.upenn.edu} 
  }
\address{Department of Physics, University of Pennsylvania, Philadelphia,
PA 19104-6396}
\author{Niels R. Walet  \footnote{ mccsnrw@afs.mcc.ac.uk}}
\address{
Department of Physics, University of Manchester Institute of Science and Technology,
P. O. Box 88, 
Manchester, M60 1QD, United Kingdom}

\date{\today}

\maketitle
\begin{abstract}
In two previous papers, the Kerman-Klein-D\"onau-Frauendorf (KKDF)
model was used to study rotational bands of odd deformed nuclei.  Here
we describe backbending for odd nuclei using the same model.  The
backbending in the neighboring even nuclei is described by a
phenomenological two band model, and this core is then coupled to
a large single-particle space, as in our previous work. The results
obtained for energies and $M1$ transition rates are compared with 
experimental data for $^{165}$Lu and for energies alone to the 
experimental data for $^{179}$W. For the case of  $^{165}$Lu 
comparison is also made with previous theoretical work.
\end{abstract}
\section{Introduction}

\label{section:BackBending}
In two previous applications \cite{pavlos:1,pavlos:2}, the Kerman-Klein- D\"onau-Frauendorf
(KKDF) model was used to study rotational bands of selected odd
deformed nuclei.  In both applications the system was described by an
effective interaction which includes a monopole pairing and a
quadrupole-quadrupole interaction.  In the first application a large
single-particle space was coupled to the ground-state band of the
neighboring nuclei, whereas in the second application the same large
single-particle space was not only coupled to the ground-state band
but also to some of the excited bands of the core.

As a second class of applications, we want to describe backbending for
odd nuclei using the KKDF model.  As has been known for more than two
decades, deformed nuclei commonly show a rotational anomaly (known as
backbending), where the energetically favored, or yrast, collective
band undergoes an abrupt increase in its moment of inertia (as a
function of frequency, for example).  The generally accepted interpretation is that
backbending of an even nucleus occurs when two neutrons (or protons)
with high $j$ break their pairing bond and rotationally align
perpendicular to the symmetry axis \cite{Stephens}. The
rotational-aligned sequence of states is called the s-band.  In this
application, instead of following a microscopic approach to
backbending that incorporates this physics, we have chosen to take a
purely phenomenological approach using a model of two coupled
bands. These will then be coupled to the extra odd particle in order
to achieve a description of the corresponding phenomena in the
neighboring odd nuclei. The reason for utilizing a phenomenological 
description of the even cores is that this provides a more 
accurate fit to the data than existing microscopic calculations.

In the next section we will present the description of the cores,
including the phenomenological model and the results. In
Sec.~\ref{sec:W179} the results of the energy calculations for
$^{179}$W will be presented, and in Sec.~\ref{sec:Lu165} the results
for both energies and $M1$ transitions will be given for
$^{165}$Lu. There is a brief concluding section.

\section{Core Phenomenology} 
In this section we develop a phenomenological description of the
backbending phenomenon for the even neighboring nuclei.  It is not our
purpose to build a complete and sophisticated model but to reproduce
the energy levels and the values of the quadrupole matrix elements as
accurately as possible, because the results of these calculations will
be used as input to the calculations for the odd nuclei. The purpose of the 
model is simply to provide input data to our theory when experimental
values are not sufficiently abundant.

Consider two bands, the ground-state band and an excited band that
``cross'' at a certain angular momentum $I$. From experimental
observations we know that if such a situation occurs, the bands repel
each other (avoided crossing).  From the mixture of the bands, we
can conclude that the Hamiltonian of the system is in general not
axially symmetric, and as a result, the projection $K$ is not a good
quantum number.  We write the Hamiltonian of such a system as
\begin{equation} 
  H = H_{0}+ H_{1},
\end{equation}
where $H_{0}$ is the axially symmetric part of Hamiltonian and $H_{1}$
is the $K$ quantum number non-conserving Hamiltonian which gives rise
to the avoided crossing.  The unperturbed Hamiltonian, $H_{0}$,
describes the ground-state rotational band and one excited band.  The
excitations in such bands can have the form of the simple rigid rotor, or
of the variable moment of inertia model (VMI model) \cite{VMI1,VMI2},
or even more general form (see below). For the uncoupled systems, we
take the angular momentum $I$, its projection $M$, and the projection
on the body fixed axis $K$ as good quantum numbers. Therefore, we
denote the eigenstates of the unperturbated system as $ | IMK \rangle
$.  Expressed in the unperturbated basis, the Hamiltonian for a given
angular momentum will take the following form,
\begin{equation} 
  H= \frac{\hbar^{2}}{2}\left[
  \begin{array}{cc} 
    f_{1}(  {\hat I}^{2}  ) & 0 
    \\ 
    0 & E_{o}+ f_{2}(  {\hat I}^{2})
  \end{array}
\right] + \left[
\begin{array}{cc}
  0 & C( {\hat I}) \\ C( {\hat I}) & 0 \\
\end{array} 
\right],
\end{equation}
where $ {\hat I} = I(I+1) $ and $E_{o}$ is the bandhead energy.
 The perturbation has, for the moment, a general
 ${\hat I}$ dependence.  The functions $f_{i}({\hat I}) $ which
 describe the angular momentum dependence of the uncoupled band $i$
 ($i=1,2$) are often represented as a polynomial in $I(I+1)$,
\begin{equation}
f_{i} (  {\hat I}^{2} ) = \frac{ I(I+1) }{ {\cal I}^{0}_{i}
     } + \frac{ (I(I+1))^{2} }{ {\cal I}^{1}_{i} } + \cdots,
\end{equation}
with expansion coefficients ${\cal I}^{0}$, ${\cal I}^{1}$.  This
conventional expansion has poor convergence properties at large $I$.  
A better expansion of energy as a function of angular momentum 
is given by the {\em variable moment of inertia} model (VMI) \cite{VMI}  in which the moment of 
inertia is a function of angular momentum
\begin{equation}
E(I, {\cal I})= \frac{ I(I+1)}{ 2 {\cal I}(I)} + \frac{1}{2} C\, \left( {\cal I}(I) -{\cal I}_{0}\right)^{2},
\end{equation}
with two parameters  $C$ and ${\cal I}_{0}$. The
variable moment of inertia ${\cal I}(I)$ is
determined through use of the variational condition 
\begin{equation}
\left. \frac{ d E(I)}{d {\cal I}(I)} \right| _{I}=0.
\end{equation}
This
model can be shown to be equivalent \cite{Klein_VMI} to the two-term approximation
to the Harris (cranking) formula \cite{Harris}.  The latter
is an expansion of the energy in powers of the ``rotational frequency''
$\omega$, given by
\begin{equation} 
     E(\omega) =  {\cal A}^{0} \, \omega^{2} + 
  {\cal A}^{1}  \, \omega^{4}+  {\cal A}^{2} \, \omega^{6} + \cdots ~,
\label{eq:wexp}  
\end{equation} 
where ${\cal A}^{0}$ , ${\cal A}^{1}$ etc are parameters that are
fitted to experimental values.  Since the concept of continuous
angular frequency is not well defined for quantum mechanical systems,
the definition used is derived from the corresponding classical
definition. In classical mechanics we have,
\begin{equation} 
 \omega(I) = \frac{ dE(I)}{ dI},
\end{equation}  
where $E(I)$ is the energy at angular momentum $I$.  In quantum
mechanics we take a cue from this definition and choose the discrete
definition
\begin{equation} 
\omega(I) = \frac{ E(I_{+1}) - E(I) }{ {\hat I_{+1}} - {\hat I}}
,
\end{equation} 
where $I_{+1}$ is the next angular momentum value that the system can
take. In our calculations we choose to use the Harris formula to
the third order.

At this point we have to specify what form the coupling will take. The
exact form of the inter-band interaction is not known. We have chosen
to perform the calculations using the standard band coupling formalism
\cite{ref:B_M_book}. Since the ground state has $K=0$, the possible cases are limited by the 
values of $K$ chosen for the excited band. In the following we allow $K=0,1~\rm and~ 2$.
\begin{eqnarray}
{\rm For} ~ K=0 \rightarrow K=1: & & \hspace{1cm}
    C(I) = C \sqrt{I(I+1)},
\\   
{\rm For}~ K=0 \rightarrow K=0:  & &\hspace{1cm}
    C(I) = C I(I+1),
\\
{\rm Constant}~{\rm interaction}: & &  \hspace{1cm}
    C(I) = C. 
\\
{\rm For} ~ K=0 \rightarrow K=2: & &\hspace{1.0cm}  
C(I) = C (I+1) (I+2)(I-1).
\end{eqnarray}
%
%{\rm   For}~  K=2 \rightarrow K=1: & &  \hspace{1cm}
%    C(I) = C \sqrt{ (I-1) (I+2) },  
%\\
%\\ 
%{\rm For}~ K=2 \rightarrow K=2: & & \hspace{1cm} 
%C(I) = C ( I (I+1)-4),
%\end{eqnarray}
The actual angular momentum dependence of the coupling
term is not decisive for the following reason: The 
mixture of two uncoupled band states
with the consequent distortion of the shapes of the energies as functions of the 
angular momentum depends on the strength and
the form of the coupling term as well as on the energy difference of
the unperturbed states.  The bands will be close to each other (in the
Energy vrs. $J$ plot) only for a single value of angular momentum, and
therefore the mixture will be insignificant for all other values of
angular momentum.  As soon as the strength of the coupling term is
fitted at this angular momentum, the actual functional dependence of
the coupling strength becomes irrelevant. For the actual calculations
we tried different forms of the coupling and we state in Table
\ref{table:back1} which ones yield the best fit for the nucleus under
study.
 
At this point we have a complete phenomenological description for the
backbending phenomena. If we use to the expansion
Eq.~(\ref{eq:wexp}) up to the third order, we then have eight free
parameters to fit, namely the six coefficients for the two uncoupled
bands, $ {\cal A}^{0}_{1}$, $ {\cal A}^{1}_{1}$, $ {\cal A}^{2}_{2}$,
$ {\cal A}^{0}_{2}$, $ {\cal A}^{1}_{2}$ and $ {\cal A}^{2}_{2}$, the
band-head energy $E_{0}$, and finally the coupling $C$. This is not a
difficult task, and we will describe the details and show the results in the
following section.

To calculate the $BE(2)$'s we have to express the matrix elements of
the quadrupole operator as a function of the matrix elements in the
uncoupled system.  To be more specific with the notation, states of
the unperturbated Hamiltonian that belong to the ground-state band will be
denoted by $ | IM0 \rangle $ and states that belong the excited band $
| IMK \rangle $. For the coupled system states have good angular
momentum $I$, its projection $M$, but $K$ is not a good quantum number
any more.  We denote states of the complete system as $ | IM1 \rangle
$ and $ | IM2 \rangle $.  After diagonalizing the Hamiltonian, we can
express the eigenstates of the full system as a linear combination of
the unperturbed states
\begin{eqnarray}
|IM1 \rangle  &=& \lambda_{0}(I) | IM0 \rangle  + \lambda_{1}(I)
| IMK \rangle ,   \nonumber \\
|IM2 \rangle  &=& \mu_{0}(I) | IM0 \rangle  + \mu_{1}(I)
| IMK \rangle  
,
\end{eqnarray}
where $\lambda_{0}(I)$, $\lambda_{1}(I)$, $\mu_{0}(I)$ and
$\mu_{1}(I)$ will be provided from the diagonalization of the
Hamiltonian.  Consequently we can express the matrix elements of the
quadrupole operator, in the coupled case, as a function of the matrix
elements of the unperturbated case,
\begin{eqnarray} 
\label{eq:ME}
 \left< IM1 | {\hat Q} | I'M'1 \right> & =&
      \lambda_{0}(I) \lambda_{0}(I')     \left< IM0 | {\hat Q} | I'M'0 \right> 
      + \lambda_{1}(I) \lambda_{1}(I')     \left< IMK | {\hat Q} | I'M'K \right> 
      \nonumber \\ 
      & +& \lambda_{0}(I)  \lambda_{1}(I') \left< IM0 | {\hat Q} | I'M'K \right> 
      + \lambda_{1}(I)  \lambda_{0}(I') \left< IMK | {\hat Q} | I'M'0 \right> \nonumber ,
     \\ 
 \left< IM1 | {\hat Q}_{\mu} | I'M'2 \right> & =&
      \lambda_{0}(I) \mu_{0}(I')     \left< IM0 | {\hat Q} | I'M'0 \right> 
      + \lambda_{1}(I) \mu_{1}(I')     \left< IMK | {\hat Q} | I'M'K \right> 
    \nonumber  \\ 
      & +& \lambda_{0}(I)  \mu_{1}(I') \left< IM0 | {\hat Q} | I'M'K \right> 
      + \lambda_{1}(I)  \mu_{0}(I') \left< IMK | {\hat Q}| I'M'0 \right> \nonumber ,
     \\ 
     \left< IM2 | {\hat Q}_{\mu} | I'M'2 \right> & =&
      \mu_{0}(I) \mu_{0}(I')     \left< IM0 | {\hat Q} | I'M'0 \right> 
      + \mu_{1}(I) \mu_{1}(I')     \left< IMK | {\hat Q} | I'M'K \right> 
     \nonumber  \\ 
      & +& \mu_{0}(I)  \mu_{1}(I') \left< IM0 | {\hat Q} | I'M'K \right> 
      + \mu_{1}(I)  \mu_{0}(I') \left< IMK | {\hat Q} | I'M'0 \right> \nonumber .
     \nonumber  \\ 
 & & 
\end{eqnarray}
The reduced matrix elements of the quadrupole operator $ {\hat Q} $
for the uncoupled case are taken from the phenomenological
Bohr-Mottelson model,
\begin{eqnarray} 
\label{eq:BM}
\left< I0 \parallel {\hat Q} \parallel I'K \right> &=& \sigma_{K'}~
          q^{{\rm band1} \rightarrow {\rm band2}} \sqrt{2I+1}
          \sqrt{2I'+1} \left(
     \begin{array}{ccc}
          I & I' & 2 \\ 0 & K & -K \\
     \end{array}
     \right), \\ \sigma_{K'} &=& \left \{
   \begin{array}{c} 
     \sqrt{2} \hspace{1cm} K' \neq 0 \\ 1 \hspace{1.2cm} K' =0
     \end{array} \right.  \nonumber ,
\end{eqnarray}
where $q^{{\rm band1} \rightarrow {\rm band2}}$ are the intrinsic
quadrupole moments of any inter or intra-band transitions.
Consequently we have three more parameters to fit, namely the three
intrinsic quadrupole moments. From measured $BE(2)$'s we fix these three
parameters and then calculate the rest of the $BE(2)$'s that are needed in
the calculation but not available experimentally.

\begin{figure}
\centerline{\epsfysize=8cm \epsffile{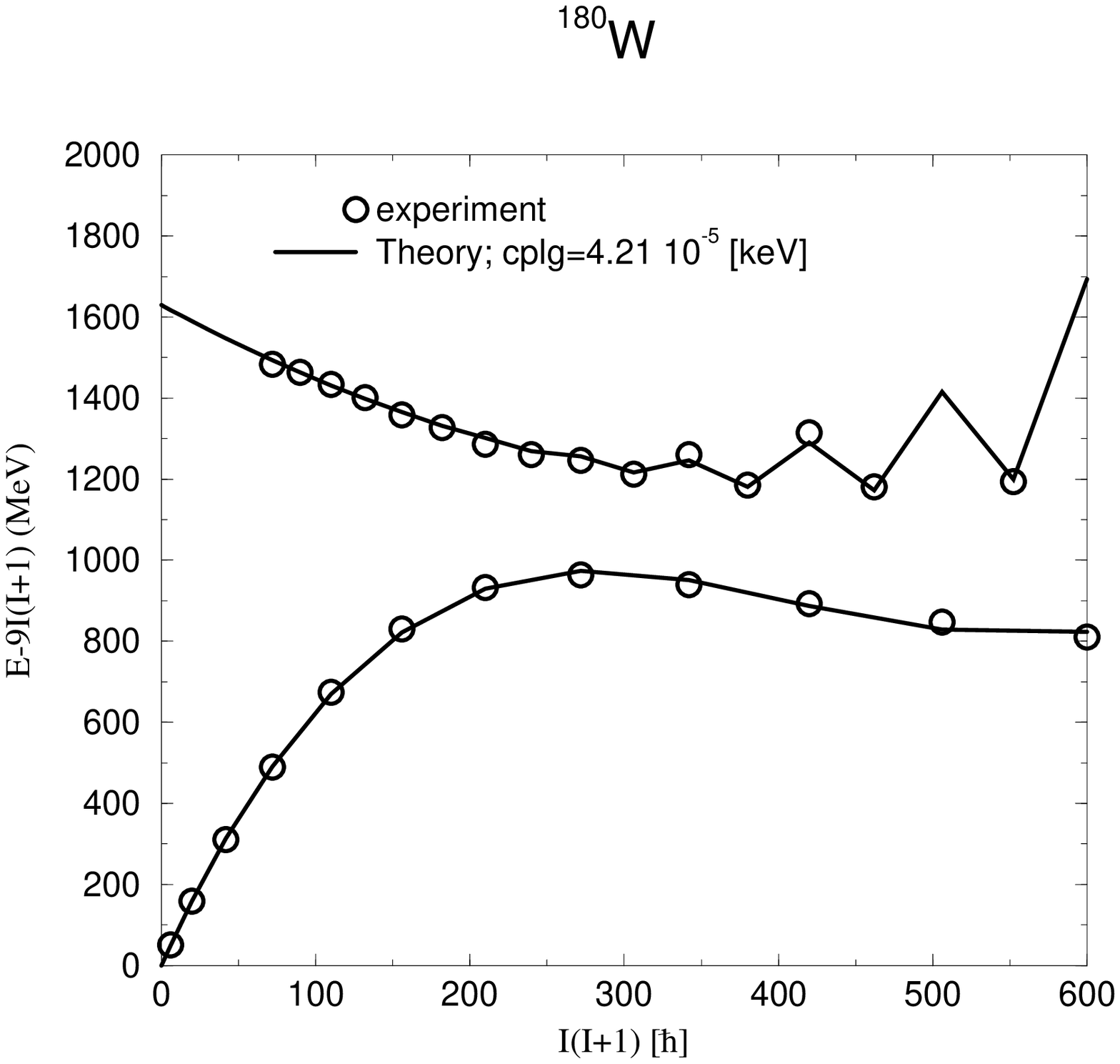}}
\caption{ \label{fig:W180_lev}
Fit of the ground-state band and the first excited band of $^{180}$W
to a phenomenological two-band model.  Solid lines represent the
theoretical values with coupling equal to 4.3~$10^{-5}$keV.  The
experimental values are represented by circles. }
\end{figure}

For the first application, $^{179}$W, we need the excitation energies
and the quadrupole matrix elements of the neighboring even-even nuclei
$^{178,180}$W.  Since we assume particle-number non-conservation the
difference between the two neighboring nuclei should be small, and
since there are not enough experimental data for $^{178}$W, we 
did the calculations for $^{180}$W only.  In the special case where the
excited band has odd values of angular momentum the fitting can be
simplified further.  These odd angular momentum states do not
interact with the states of the ground state band, and since they
belong to the same band, the parameters ${\cal A}$ of this band can
be calculated independently. First, we fit the parameters $ {\cal
A}_{\rm{exc}}^{0,1,2}$ for the states with odd value angular momentum, and
then we fit the other 5 parameters. For example the excited band of
$^{180}$W (see Fig.~(\ref{fig:W180_lev})) has states with odd value
$I$. Therefore, we fitted the upper band first and when the $ {\cal
A}_{\rm{exc}}^{0,1,2}$ were found, we fitted the rest.  The results are
shown in Table (\ref{table:back1}) along with Fig.~(\ref{fig:W180_lev}).

\begin{figure}
\centerline{\epsfysize=8cm \epsffile{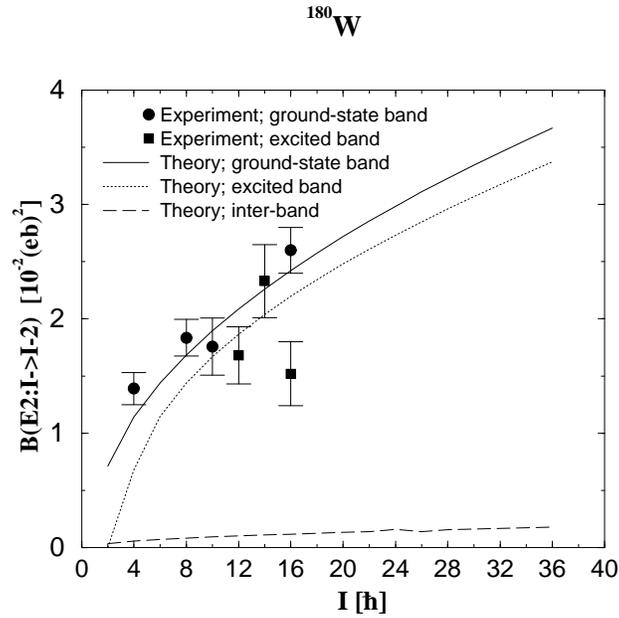}}
\caption{ \label{fig:W180_be2}
Experimental and theoretical $BE(2)$ values for the yrast band 
and first excited band. }
\end{figure}

For the $BE(2)$'s we utilize the few experimental data
available. As stated above, there are three parameters to fit, $q_{0}$, $q_{K}$ and
$q_{K0}$. With these parameters fixed, we calculated all transitions
that will be needed for the calculations of the odd nuclei. The
results of the fit are given in Fig.~(\ref{fig:W180_be2}) and
Table~(\ref{table:back1}).

\begin{figure}
\centerline{\epsfysize=8cm \epsffile{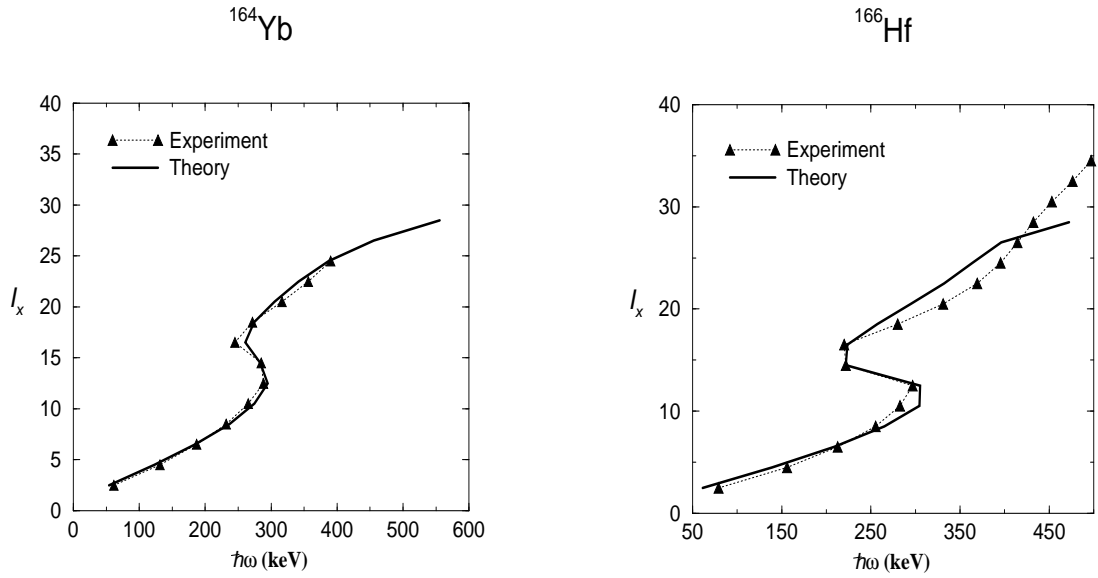}}
\caption{ \label{fig:Hf_Yb}  Energy levels of $^{164}$Yb (left) and 
$^{166}$Hf (right). Both experiment and theory are shown.}
\end{figure}

\begin{figure}
\centerline{\epsfysize=8cm \epsffile{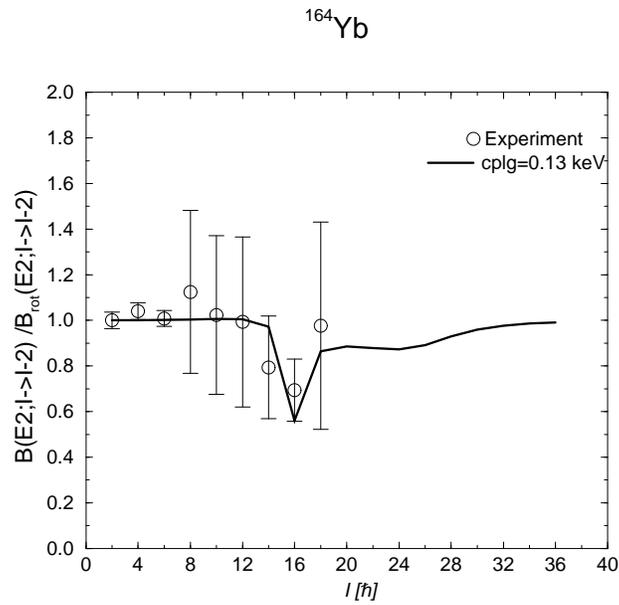}}
\caption{ \label{fig:Yb164_be2}
 $BE(2)$ values from experiment (circles) and theoretical fit (solid line) of  $^{164}$Yb. }
\end{figure}

As a second application we considered the odd proton nucleus
 $^{165}$Lu.  The neighboring even nuclei are $^{164}$Yb and
 $^{166}$Hf. There are enough experimental values for both these
 nuclei, and therefore we have fitted both nuclei and taken the
 average values of the excitation energies and quadrupole matrix
 elements for application to the odd nucleus. The calculation was done
 as described above for $^{180}$W with one modification however. We
 realized that the fitting of the $BE(2)$ values is more sensitive to
 the strength of the coupling $C(I)$ than are the excitation energies.
 Therefore, we first do a preliminary fit of the excitations energies.
 We then adjust the wavefunctions found from these calculations to fit
 the $BE(2)$ values by allowing $C$ to vary once more, in addition to
 the necessary choice of intrinsic quadrupole moments.  The results
 are shown in Fig.~(\ref{fig:Yb164_be2}) and in Table
 (\ref{table:back1}). Then the excitation energies are re-calculated
 with an already fixed coupling strength $C$, and we find new values
 for the ${\cal A}$'s.  The results of final fitting of the excitation
 energies are shown in Fig.~(\ref{fig:Hf_Yb}) and in
 Table~(\ref{table:back1}).

\section{Calculations and results for $^{179}$W}
\label{sec:W179}
The equations of motion (EOM), as well as the normalization
conditions, for the case of backbending are not formally different
from those presented in the two previous applications
\cite{pavlos:1} and therefore will not be repeated.  
In contrast to these applications, the
quadrupole field $\Gamma$ will not take one of the simplified forms
of the Bohr-Mottelson theory, such as Eq.~(\ref{eq:BM}).
Instead, it will take values that are
determined from the calculation described in the previous section, culminating in 
Eq.~(\ref{eq:ME}). The form of $\Gamma$ is given by,
\begin{equation}
     \Gamma_{aIn,cI'n';J} = 
     -\frac{1}{2} \, \kappa (-)^{j_{c}+I+J}
     {\setlength{\arraycolsep}{2pt} \left\{\begin{array}{ccc}
       {\scriptstyle  j_{a}}&{\scriptstyle j_{c}}&{\scriptstyle 2}
       \\ 
       {\scriptstyle I'}&{\scriptstyle I}&{\scriptstyle J}
\end{array}\right\}}
\left< In \| Q \| I'n' \right>,
\end{equation}
where $n=1,2$ , $n'=1,2$, and the quadrupole matrix elements, $\left< In \| Q \| I'n'\right>$,
 were calculated in the previous section. Notice that $\Gamma$ depends
 on the total angular momentum $J$ in contrast to the two previous
 cases. In the previous calculations, because we assumed axial
 symmetry, $\Gamma$ was independent of J.  In this application the
 symmetry is lost, with implications that will be discussed later.

Similarly to the previous two examples, we used a large
single-particle space (including all states from 5 major shells). The
energies and matrix elements of these single-particle levels were
calculated using the Woods-Saxon potential.  The core excitations
$\omega_{I}$ were taken from calculations for the core described in
the previous section.  At this point all input parameters are fixed
and the only thing remaining to be done is to solve the eigenvalue
problem given by the EOM. In the same way as before, the strength of
the quadrupole field is treated as a free parameter and the values of
the single-particle energies found from Woods-Saxon calculations are
allowed to vary by $\pm 5 \%$.

Two remaining technical problems should be discussed before we present
the results of the calculations. The first difficulty in solving the
EOM is that the set of solutions is over-complete by a factor of two,
as has previously been discussed.  This is a consequence of the fact
that the basis states form an over-complete (and, therefore,
non-orthogonal set).  Consequently half of the states found by solving
the EOM are not physical and have to be removed.  A technique has been
developed to do so and was presented in \cite{pavlos:1}.  In \cite{pavlos:2}, an
extension of the technique to the case of avoided crossings, was discussed.
In this paper, we summarize only the essential
points. The Hamiltonian is first decomposed into anti-symmetric and
symmetric parts with respect to particle-hole conjugation. If
only the anti-symmetric part is diagonalized, then for every positive
energy eigenvalue there is a negative partner. From the BCS theory we
know that the positive eigenvalues are the physical solutions and the
negative solutions the non-physical ones. Then the symmetric part was
turned on ``slowly'' and at every step the physical solutions were
identified using the projection operator for the physical space
 built from the wavefunctions of
the previous step. The 
projection operator  has an eigenvalue 1 for all real states and
0 for the non-physical states. Since this operator is
constructed from wavefunctions of the previous step we expect that it
will give an expectation value close to 1 for physical and close to 0 for the
non-physical states of the current step.  This procedure continues until the symmetric
part is turned on fully. In \cite{pavlos:2}, we explained the case where
physical and non-physical states of the same angular momentum come
very close to each other, and consequently the corresponding
wavefunctions change rapidly. In this instance the projection operator
method is not valid unless the steps taken are extremely small,
making the actual calculation numerically very slow.  In this case we
utilized the fact that states of the same angular momentum can not
actually cross.  As stated previously, the two bands involving the
near crossing interchange the characters of their wavefunctions as
they pass near each other. Since the projection operator method is
basically a comparison of wavefunctions, 
the form of this operator just prior to the crossing  would
classify a physical state as unphysical and vice versa when used in
coarse steps involving the near-crossing.  Therefore when this
occurs, we adjust the projection operator appropriately in order to make
the correct identification in future steps.  The problem in now
reduced to determining if there is a crossing. This was done by
comparing the energy differences between two states at two consecutive
steps.  This method worked extremely well both in the example of the
previous work \cite{pavlos:1,pavlos:2} and in the present case.

The second technical problem is the classification of states into
different bands. In the first two applications where $K$ was a good
quantum number, the identification of bands was done based on the $K$
value of the band. The technique was explained in \cite{pavlos:1} and again
we repeat only essential aspects. In the case that the core is
approximated as a rigid rotor, and the formulas of the geometrical
model are valid, it turns out that the quadrupole field is independent
of the total angular momentum $J$.  This is because of the assumed
axial symmetry.  As a result when only the antisymmetric part of the
Hamiltonian is considered the Hamiltonian has no $J$ dependence. It
follows that for a given $K$, i.e. a given band, all eigenstates with
different $J$ are degenerate.  We first solve our equation for a
minimum possible angular momentum, $J=\frac{1}{2}$. This identifies
the $K=\frac{1}{2}$ bands. At $J=\frac{3}{2}$, we find a set of
energies equal to those previously found and possibly additional
solutions identified as the band heads for $K=\frac{3}{2}$. Continuing
in this way we assemble all solutions into a series of flat bands with
identified $K$ values.  

In the case that backbending occurs, even when
the core excitations are neglected (symmetric part of H set to zero)
the remaining Hamiltonian is J dependent.  Therefore, we can not use
the method described above. We can however, take the additional steps
of turning off the coupling between the two core bands to return to
the axially symmetric case. Then the previous procedure can be used,
but there is the additional complication of having to turn on both the
symmetric part and the coupling between the two core bands. We can do
that in any order, i.e. we can first turn on the coupling between
the core bands and then the core excitation (symmetric part of the
Hamiltonian) or the other way around.  We choose to do the former.
First we turn on the band coupling slowly, and as a result we break the
axial symmetry. At every step we compare the wavefunctions at two
consecutive steps by calculating the overlap integral. It turns out
that even though the $J$ degeneracy is lifted the wavefunctions of the
axial and non-axial cases (not including the core excitations) are
very similar.  Only for a few eigenstates was the mixture between
states belonging to different $K$ bands so big we had to turn on
the coupling slowly. In most cases large steps were sufficient.

The first application made was to the nucleus $^{179}$W. Recent
observations \cite{Walker1,Walker2} have been interpreted as
showing that for this odd-neutron nucleus, an alignment to an axis
intermediate between the deformation axis and the collective
rotational axis (Fermi alignment \cite{Frauend}), gives rise to a
backbending at $J \sim \frac{31}{2}$. (Frauendorf \cite{Frauend} has
called the associated rotational structure a t-band, since the
cranking-model description requires a {\it tilting} of the cranking
axis away from the principal axes of the intrinsic prolate spheroidal
shape.) Nevertheless we have chosen to try the standard two-band
model.

\begin{figure}
\centerline{\epsfysize=8cm \epsffile{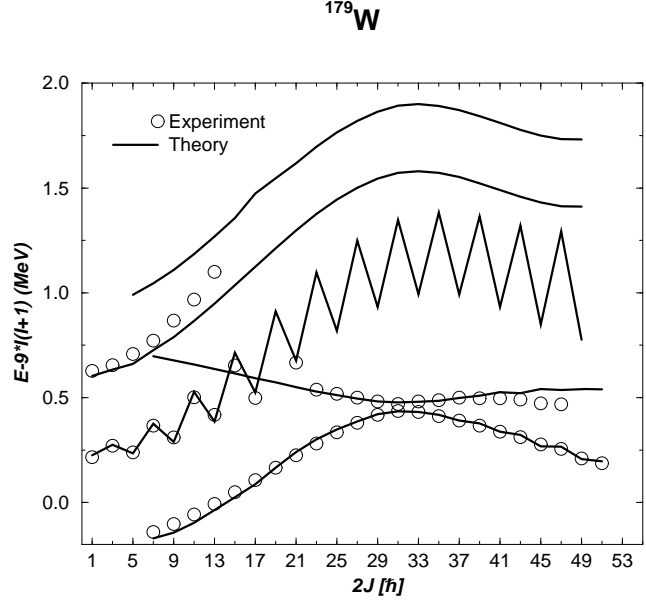}}
\caption{ \label{fig:W179}
Comparison of theoretical results to observed energy levels of $^{179}$W.  }
\end{figure}

Energy levels of $^{179}$W calculated from the present work are
presented in Fig.~(\ref{fig:W179}).  As can be seen from the figure,
the yrast band ($K=7/2$) is reproduced with high accuracy. The same
can be said of the first excited band ($K=7/2$ or tilted-band). Then
follow two $K=1/2$ bands which agree very well with the theory.  The
most striking feature is the exact reproduction of the staggering
nature of the first $K=1/2$ band. At about 1 MeV above the ground
state, theory predicts a $K=5/2$ band which experimentally is not
observed.  We believe that this a weakness of the fitting routine for
the following reasons. The relative band-head energies were fitted,
using a standard fitting routine, to the observed values by varying
the strength of the quadrupole field, $\kappa $. The fitting was done
by minimizing the $\chi^{2}$. In the absence of an experimental value
and the presence of a a theoretical prediction the contribution to the
$\chi^{2}$ is 0.  In other words for the fitting routine (at least for
the one we used), if the theory predicts a band at low energy that is
not observed experimentally it does not increase the $\chi^{2}$. We
are currently working to build a better fitting routine which will be
able to avoid such a problem. The result shown in Fig.~\ref{fig:W179}
already represents some improvement over the fitting procedure used
initially.

\section{Results for $^{165}$Lu}
\label{sec:Lu165}
\begin{figure}
\centerline{\epsfysize=8cm \epsffile{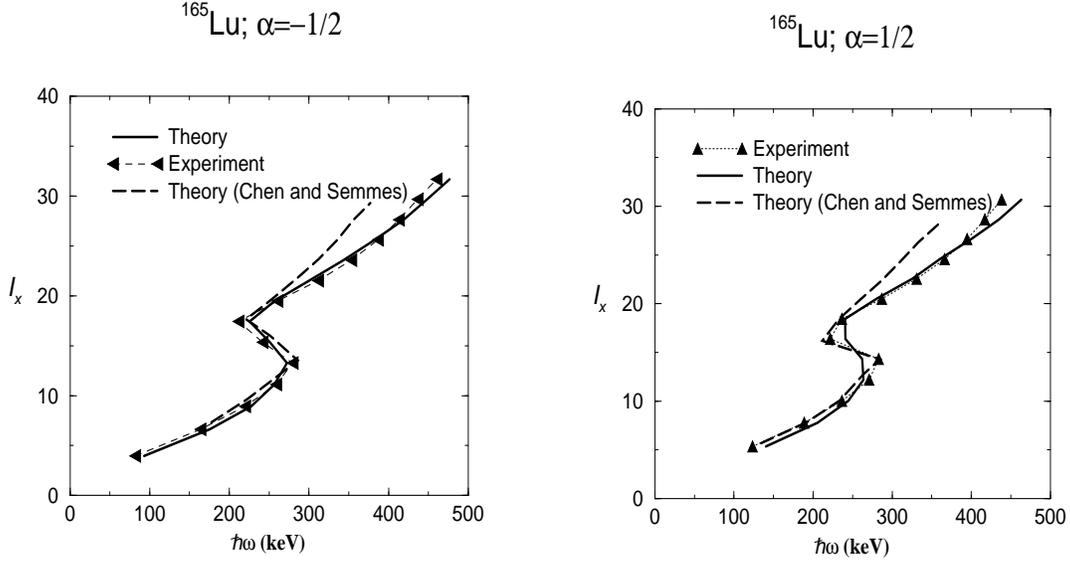}}
\caption{ \label{fig:Lu165}
 Comparison of the observed yrast band of  $^{165}$Lu with theory. 
On the left are the negative signature states ($\alpha=-\frac{1}{2}$) 
and on the right are the positive signature states  ($\alpha=\frac{1}{2}$).   }
\end{figure}

The second backbending application we tried was to the proton spectra of
 $^{165}$Lu.  This nucleus was studied previously by methods related to
those of this paper by Chen {\it et al} \cite{ref:Chen}.  Since it is
generally accepted that the s-band in an even nucleus is formed by the
decoupling of a pair of quasiparticles from the ground-state band,
these authors, among others, decided that it was necessary to use a
semi-microscopic description of the backbending in the neighboring
even nuclei. However, their fit to the spectra above the backbend
leaves something to desired.  For this reason, we have chosen to
separate the problem of fitting the backbend in the even nuclei from
that of fitting the corresponding spectra in the odd nuclei by using a
phenomenological two-band mixing model for the even nuclei, which
allows, simultaneously an accurate fit to the the spectrum of the
crossing bands and to all observed B(E2) values (see previous section).
In Fig.~(\ref{fig:Lu165}), we compare our fit for both negative and positive 
signature ($\alpha = -\frac{1}{2}$,  $\alpha = \frac{1}{2}$) crossed 
bands in $^{165}$Lu with those of Chen. The closeness of our fit compared
to that of the previous authors appears simply to reflect corresponding 
deviations in the spectra of the even neighbors. 

\begin{figure}[h]
\centerline{\epsfysize=8cm \epsffile{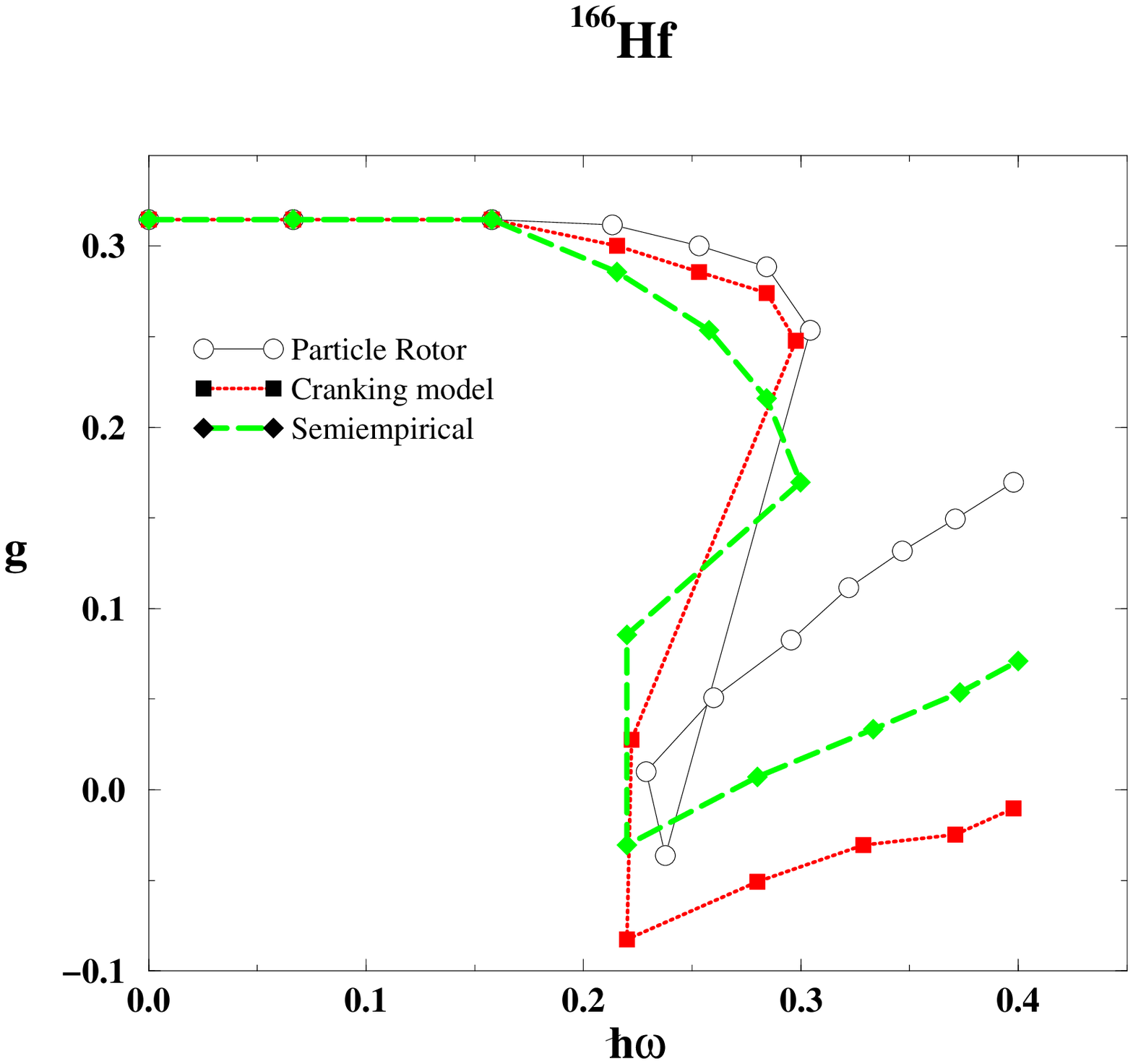}}
\caption[]{ \label{fig:Hf166_m1}
 Calculated $g$ factors for the yrast states of $^{166}$Hf. Squares are 
calculations from the cranking model, the circles are from the two-quasiparticle
plus rotor model, and the triangles are from the 
\protect{\cite{Frau81}}. 
}
\end{figure}

We next study magnetic transitions of the yrast band and compare the
result to experiment and other theories.  In \cite{pavlos:1} we have
presented the equations for the $M1$ matrix element and calculations
for the $B(M1)$ transitions of the ground state band of
$^{155,157}$Gd.  To calculate $B(M1)$ rates, we must first evaluate
the matrix element of the $M1$ operator between states of the even
neighboring cores $\langle I \| M \| I \rangle$ and the matrix
elements of the same operator between single particle states
($m_{ac}$).  In the case of $\langle I \| M \| I \rangle$, and because
of the limitation on the availability of the experimental data, we
have chosen to use a phenomenological model for the value
of $g$, that is derived in \cite{Frau81}, namely
\begin{equation}
  g = g_{R} + ( g_{j} -g_{R} )\, i/ I ,
\end{equation}
where $g_{R}$ refers to the ground band, $i$ is the aligned angular momentum,
and $g_{j}$ is the single particle $g$ factor for the $j$ shell in which alignment 
occurs. For the $\nu_{13/2}$ level, $g_{j}$ takes the value of $-0.20$ and $g_{R}=0.31$.
In Fig.~\ref{fig:Hf166_m1} we show the $g$ factors for the yrast band of $^{166}$Hf
obtained from the cranking model, the two-quasiparticle plus rotor model, and 
from the simple formula shown above.
For the single particle matrix elements we used a simple formula from
\cite{ref:Ring_Shung}, Eq.~(44) of \cite{pavlos:1}.

\begin{figure}[ht]
\centerline{\epsfysize=14cm \epsffile{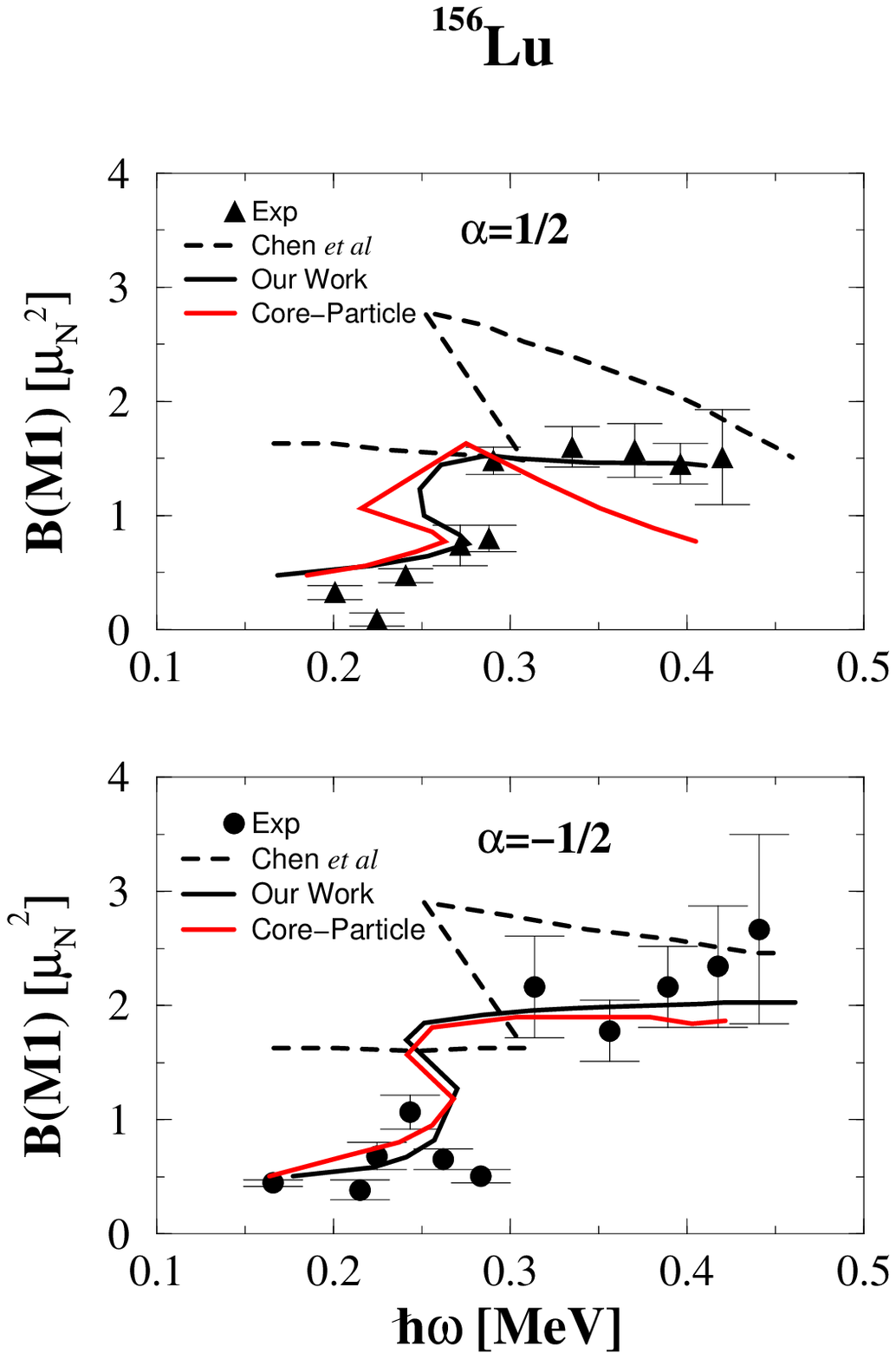}}
\caption[]{ \label{fig:Lu165_m1}
 $B(m1)$ transitions rates in $\mu_{N}^{2}$. On the top are the
positive signature states ($\alpha=-\frac{1}{2}$) and on the bottom are
the negative signature states ($\alpha=\frac{1}{2}$).  
Experimental data
 \cite{Jons84} 
are solid symbols, strong coupling 
model is the chain lines, our work is represented with the solid lines
and the work of Chen {\em et al} 
\protect{\cite{ref:Chen}}
 is the dashed lines.}
\end{figure}

Results from different calculations of $B(M1)$ rates of the yrast band
 of $^{165}$Lu are summarized in Fig.~\ref{fig:Lu165_m1}.  Figure
 \ref{fig:Lu165_m1} shows experimental values from \cite{Jons84}, the
 result from the cranking calculation of \cite{ref:Chen}, a
 core-particle coupling model also presented in \cite{ref:Chen} and
 finally our calculations.  Our work reproduces the experimental values
 quite well and clearly better than the other works.

\section{Conclusions}
In this paper we have presented yet another distinct application of the KKDF
model.  Calculations for $^{165}$Lu and $^{179}$W were performed
starting from a phenomenological description of the neighboring even
nuclei.  The results are sufficiently satisfactory to encourage us to
consider still more sophisticated applications of the model utilized.
A fundamental approximation that all
applications so far shared was that the incorporation of
particle-number conservation into the equations has little effect on
the results, because the properties of the neighboring nuclei are so
smooth and slowly-varying with particle number. It therefore behooves
us to study a case where the incorporation of exact number
conservation is essential because we are in a transitional
region. Such an has been worked out \cite{pavlos:thesis}.

\acknowledgments
This work was supported in part by the U.S. Department of Energy 
under Grant No. 40264-5-25351.

\bibliographystyle{aip}
\bibliography{mybib}

\begin{thebibliography}{10}

\bibitem{pavlos:1}
P.~Protopapas, A.~Klein, and N.~R. Walet,
\newblock Phys. Rev. {\bf C 50}, 245 (1994).

\bibitem{pavlos:2}
P.~Protopapas, A.~Klein, and N.~R. Walet,
\newblock Phys. Rev. {\bf C}  (in press).

\bibitem{Stephens}
F.~S. Stephens,
\newblock Rev. Mod. Phys. {\bf 47}, 43 (1975).

\bibitem{VMI1}
M.~A.~J. Mariscotti et~al.,
\newblock Phys. Rev. {\bf 178}, 1864 (1969).

\bibitem{VMI2}
M.~A.~J. Mariscotti et~al.,
\newblock Phys. Rev. {\bf 178}, 1864 (1969).

\bibitem{VMI}
M.~A.~J. Mariscotti, G.~Sharff-Goldhaber, and B.~Buck,
\newblock Phys. Lett. {\bf 33B}, 333 (1970).

\bibitem{Klein_VMI}
A.~Klein, R.~M. Dreizler, and T.~K. Das,
\newblock Phys. Lett. {\bf 31B}, 333 (1970).

\bibitem{Harris}
S.~M. Harris,
\newblock Phys. Rev. {\bf B 509}, 103 (1965).

\bibitem{ref:B_M_book}
A.~Bohr and B.~R. Mottelson,
\newblock {\em Nuclear Structure Vol.2},
\newblock Benjamin, New York, 1975.

\bibitem{Walker1}
P.~M. Walker et~al.,
\newblock Phys. Rev. Lett. {\bf 67}, 433 (1991).

\bibitem{Walker2}
P.~M. Walker and et~al,
\newblock Phys. Lett. {\bf 309B}, 17 (1993).

\bibitem{Frauend}
S.~Frauendorf,
\newblock Phys. Scr. {\bf 24}, 349 (1981).

\bibitem{ref:Chen}
Y.~S. Chen, P.~B. Semmes, and G.~A. Leander,
\newblock Phys. Rev. {\bf C 34}, 1935 (1986).

\bibitem{Frau81}
S.~Frauendorf,
\newblock Physics Letters {\bf 100B}, 219 (1981).

\bibitem{ref:Ring_Shung}
P.~Ring and P.~Shuck,
\newblock {\em The Nuclear Many Body Problem},
\newblock Springer, Berlin, 1980.

\bibitem{Jons84}
S.~Jonsson et~al.,
\newblock Nucl. Phys. {\bf A422}, 397 (1984).

\bibitem{pavlos:thesis}
P.~Protopapas,
\newblock Ph D thesis, U. of Pennsylvania, 1995.

\end{thebibliography}

\begin{scriptsize}
\begin{table}[htbp]
\begin{center}
\begin{tabular}{||c|c|c|c|c|c|c|c||}
nucleus & ${\cal A}^{1}_{\rm{gs}}$ & ${\cal A}^{2}_{\rm{gs}}$ &
${\cal A}^{3}_{\rm{gs}}$ &${\cal A}^{1}_{\rm{exc}}$ & ${\cal A}^{2}_{\rm{exc}}$ &
${\cal A}^{3}_{\rm{exc}}$ & Band-Head
\\ 
 & $[ \frac{\hbar^{2}}{keV}]$ & $[ \frac{\hbar^{4}}{keV^{3}}]$&  $[ \frac{\hbar^{6}}{keV^{5}}]$ &  $[ \frac{\hbar^{2}}{keV}]$ & $[ \frac{\hbar^{4}}{keV^{3}}]$&  $[ \frac{\hbar^{6}}{keV^{5}}]$ & $[ keV]$ 
\\ \hline 
$ ^{180}W$ & 17.16 & -2.19~$10^{-2}$ & $1.95~ 10^{-5}$ & $6.45$ &$4.00 ~10^{-3}$ &
1.7~$10^{-6}$ & 1663.50
  \\ 
 $^{164}Yb$ & 18.79 & -2.56~10$^{-2}$ & $2.10~ 10^{-5}$ &  6.03 &
2.21~$10^{-3}$ & 2.10~$10^{-5}$  & 1603.34
\\
 $^{166}Hf$ & 20.55 & -2.62~10$^{-2}$ & $ 1.90 ~ 10^{-5}$ & 6.10 & -2.62~$10^{2}$
& 1.92~$10^{-5}$ &  1675.34
 \\
\end{tabular}
\end{center}
\hspace{2.2cm}
\begin{tabular}{||c|c|c|c|c|c||}
nucleus & $q_{0}$ & $q_{K}$ & $q_{0K}$ & Coupling Type & Coupling 
\\ 
 & $[b^{2}]$ &  $[b^{2}]$ & $[b^{2}]$ &   $[keV]$ &  $[keV]$  
\\ \hline
 $ ^{180}W$ & 0.711  & 0.031  & 0.66 & $C(I)=C\, (I+1)(I+2)(I-1)$ & 4.21~$10^{-5}$
\\
 $^{164}Yb$ & 0.540 & 0.021 & 0.41 & $C(I)=C \, I \,(I+1)$ & 1.30~$10^{-1}$
\\
 $^{166}Hf$ & 0.560 & 0.019 & 0.39 &  $C(I)=C \, I \,(I+1)$ & 1.37~$10^{-1}$
 \\
\end{tabular}
\caption{\label{table:back1} Values obtained for the parameters for the phenomenological
model of the even cores. The subscribe ``exc'' on ${\cal A}_{\rm{exc}}$
identifies the values of the excited bands.  }
\end{table} 
\end{scriptsize}




\end{document}